\documentclass[12pt,preprint]{aastex}

\shorttitle{Validation of Time-Distance Helioseismology}
\shortauthors{Zhao et al.}

\begin{document}
\title{Validation of Time-Distance Helioseismology by Use of Realistic
Simulations of Solar Convection}
\author{Junwei Zhao, Dali Georgobiani, Alexander G. Kosovichev}
\affil{Hansen Experimental Physics Laboratory, Stanford University,
       Stanford, CA 94305-4085}
\author{David Benson\altaffilmark{*}, Robert F. Stein}
\affil{Physics and Astronomy Department, Michigan State University,
       East Lansing, MI 48824}
\and
\author{{\AA}ke Nordlund}
\affil{Niels Bohr Institute, Copenhagen University, Juliane Maries Vej 30,
       DK-2100 K{\o}benhavn {\O}, Denmark}
\altaffiltext{*}{present address: Department of Mechanical 
Engineering, Kettering University, Flint, MI48504}

\begin{abstract}
Recent progress in realistic simulations of solar convection have
given us an unprecedented opportunity to evaluate the robustness of
solar interior structures and dynamics obtained by methods of local 
helioseismology. We present results of testing the time-distance 
method using realistic simulations.  By computing acoustic wave 
propagation time and distance relations for different depths of the
simulated data, we confirm that acoustic waves
propagate into the interior and then turn back to
the photosphere. This demonstrates that in the numerical simulations
properties of acoustic waves ($p$-modes) are similar to the solar
conditions, and that these properties can be analyzed by the
time-distance technique. For the surface gravity waves ($f$-mode), we
calculate perturbations of their travel times, caused by localized
downdrafts, and demonstrate that the spatial pattern of these
perturbations (representing so-called sensitivity kernels) is
similar to the patterns obtained from the real Sun, displaying
characteristic hyperbolic structures. We then test the time-distance
measurements and inversions by calculating acoustic travel times
from a sequence of vertical velocities at the photosphere of the
simulated data, and inferring a mean 3D flow fields by performing 
inversion based on the ray approximation. The inverted horizontal 
flow fields agree very well with the simulated data in subsurface 
areas up to 3 Mm deep, but differ in deeper areas. Due to the 
cross-talk effects between the horizontal divergence and downward flows, 
the inverted vertical velocities are significantly different from the mean
convection velocities of the simulation dataset. These initial tests
provide important validation of time-distance helioseismology
measurements of supergranular-scale convection, illustrate limitations of
this technique, and provide guidance for future improvements.
\end{abstract}


\section{Introduction}

Time-distance helioseismology, along with other helioseismology
techniques, is an important tool for investigating the solar
interior structure and dynamics. Since it was first introduced by
\citet{duv93}, this technique has been used to derive the interior
structure and flow fields of relatively small scales, such as
supergranulation and sunspots \citep[e.g.,][]{kos96, kos97, kos00,
giz00, zha01, cou04}, and also, at the global scale, such as
differential rotation and meridional flows \citep[e.g.,][]{gil97,
cho01, bec02, zha04}. These studies, together with other local
helioseismology techniques \citep[e.g.,][]{kom04, bas04, bra00},
have opened a new frontier in studies of solar subsurface dynamics.
Meanwhile, modeling efforts of time-distance helioseismology have
also been carried out to interpret the time-distance helioseismology
measurements, and provide sensitivity kernels used in inversions of
the solar interior properties \citep[e.g.,][]{giz02, jen03, bir04}.

However, despite the observational and modeling efforts, it is
difficult to evaluate the accuracy or even correctness of the local
helioseismological results because the interior of the Sun is
unaccessible to direct observations. Still, there are a couple of
approaches that help to evaluate the inverted results. One of
these is to compare the inverted solar interior structures with
models, e.g., comparing the subsurface
flow fields below sunspots \citep{zha01} with results of
sunspot models \citep{hur00}. Another approach is to
compare the flow maps obtained by different helioseismological
techniques, e.g., comparing $f$-mode time-distance measurements of
near-surface flows with measurements of flows obtained by
the ring-diagram technique \citep{hin04}. However, the first
approach is only qualitative, and although the second approach is
somewhat quantitative, there is a possibility that different
helioseismological techniques may provide similar incorrect results
because they do not account for turbulence and rapid variability of
the subsurface flows.

A convincing way to validate time-distance helioseismology is to
perform measurements and inversions on a set of realistic
large-scale numerical simulation data, which not only model the
turbulent convective motions of various scale in and beneath the
solar photosphere, but also carry acoustic oscillation signals
generated by the motions. These simulations have the following
properties: the spatial resolution is comparable to or better than
in typical helioseismological observations, the size of the
computational domain is larger than a typical supergranule,
the temporal resolution is sufficiently high to capture useful
acoustic oscillation signals, and the time duration is long enough
to extract acoustic signals with a satisfactory signal to noise
ratio. The helioseismology techniques can then be evaluated by
comparing the inverted interior results obtained from analyzing
surface acoustic oscillations with the interior structures directly
from the numerical simulation.

In this paper, we use realistic three-dimensional simulations in
solar convections by \citet{ben06}, which were based on the work of
\citet{ste00}. These simulations have
enabled us to directly evaluate the validity of time-distance
helioseismology measurements of the quiet Sun convection. In a
previous paper, \citet[][hereafter, Paper I]{geo06} have analyzed
the oscillation properties of these simulations, and found that the
power spectrum is similar to the power spectrum obtained from real
Michelson Doppler Imager (MDI) high resolution Dopplergrams. Their
analysis of time-distance diagrams also showed that the simulated
data had time-distance relations similar to those of the real
Sun. Furthermore, the near-surface $f$-mode analysis using the
simulated data gave surface structures similar to both those obtained
from local correlation tracking and those actually in the simulation. 
Thus, this set of realistic simulations of solar convection by
\citet{ben06} is quite suitable for detailed $p$-mode time-distance
studies, and allows us to evaluate the accuracy of inverted
time-distance results.

In this paper, we introduce the simulated data in \S 2. Then we
check the properties of acoustic propagation in the interior regions of the
simulated data, and make sure that acoustic signals seen at the
surface do carry information from the interior. We present these
analyses in \S 3. In \S 4, we calculate the surface sensitivity kernel
from this dataset. Then, in \S 5, we carry out
$p$-modes time-distance measurements and inversions to infer
interior flow fields, and compare our inverted results with the
corresponding simulation data. Discussion follows in \S 6.

\section{Numerical Simulation Data}

The numerical simulation data we use in this paper were results of
computation of multi-scale solar convection in the upper solar
convection zone and photosphere \citep{ben06}, using a three-dimensional 
compressible, radiative-hydrodynamic code, which employs LTE, non-gray 
radiative transfer and a realistic equation of state and opacities
\citep{ste00}.

\begin{figure}[!ht]
\epsscale{0.8} \plotone{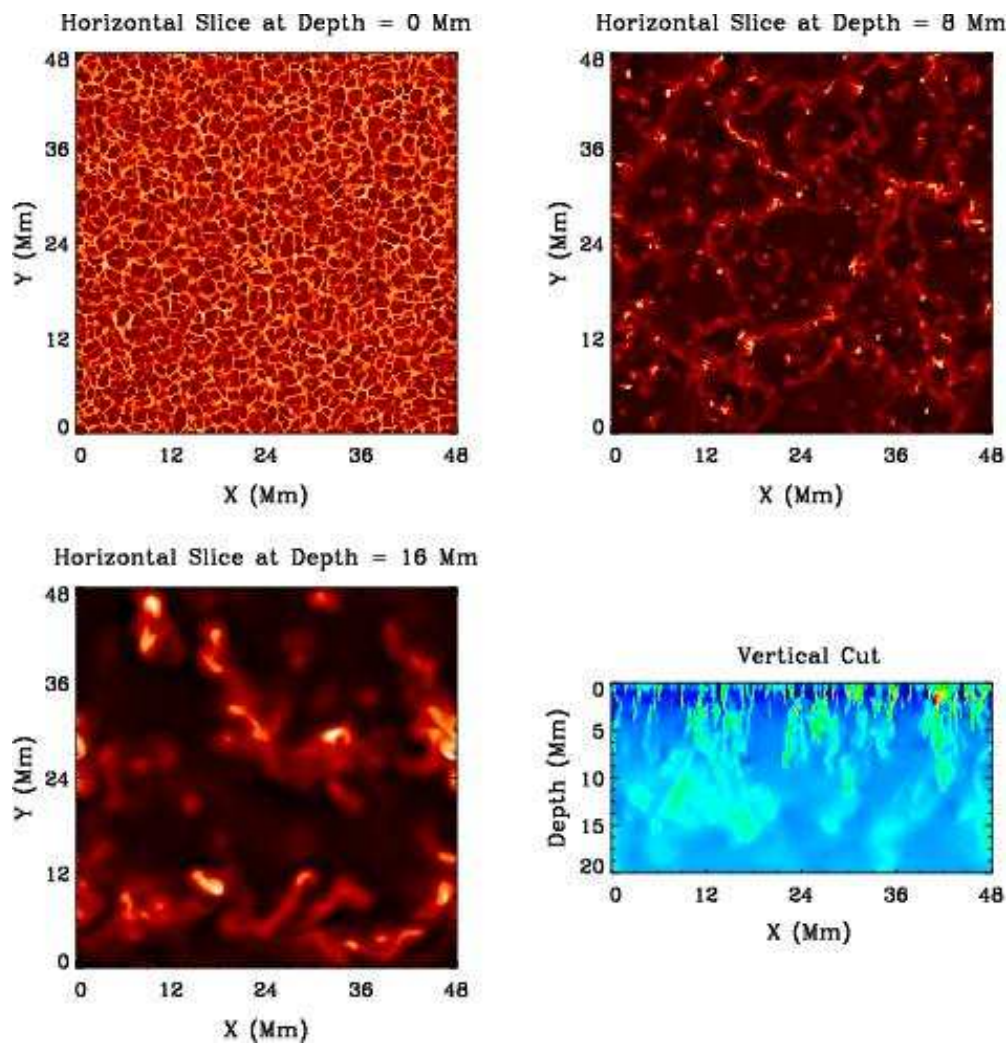} \caption{Snapshots of the vertical
component of velocity taken from the numerical simulation of solar
convection, showing three horizontal slices taken at the depths of 
0 Mm, 8 Mm, and 16 Mm, and a vertical slice, respectively. In the
vertical cut, green and yellow represents downflows, while blue and
dark represents upflows. In the horizontal slices, bright shows
downflows, and dark shows upflows.} \label{fg1}
\end{figure}

In the space domain, the simulated data span 48 Mm $\times$ 48 Mm horizontally
and 20 Mm vertically, with a horizontal spatial resolution of 96 km/pixel,
and a varying vertical spatial resolution from 12 to 75 km/pixel. In the time
domain, the data were saved every 10 seconds, but in practice, we used only
every third snapshot, i.e., every 30 seconds, because the 10-sec temporal
resolution provides an acoustic frequency range far beyond the typical
solar oscillation frequencies. The whole simulation used in this analysis
lasts 511 minutes. The acoustic $k-\omega$ diagram and the time-distance
diagram obtained from this simulated dataset in the photosphere
can be found in Paper I. In this paper and Paper I, the level of the
height of formation of the center of the $\lambda$676.78 nm Ni line 
observed by SOHO/MDI is taken as the 0 Mm level for the convenience
of description in the following text. It is 200 km above continuum 
optical depth unity.

Figure~\ref{fg1} presents a snapshot of vertical velocity 
from the simulation, showing three horizontal slices taken at 
the depth of 0, 8, and 16 Mm, and a vertical slice,
respectively. One can see small scale granular structures in the
photosphere, with downdraft lanes at the granular boundaries, and
relatively weaker upflows inside granules. Several megameters below the
photosphere, small granular structures disappear and are replaced by
larger scale structures, but with similar flow patterns, downdrafts
at boundaries and upward flows inside the structure.

\section{Propagation Properties of Acoustic Waves}

It is already clear that at the photospheric level, the simulations   
carry acoustic oscillations similar to those of the real solar 
observations, as demonstrated in Paper I. Since the goal of this 
paper is to evaluate time-distance helioseismology in the interior 
by use of $p$-modes analysis, it would be useful and necessary to
check whether the proper acoustic oscillations exist in the interior 
of the simulated convection, and whether acoustic waves propagate
inside properly.

The solar acoustic waves are excited stochastically by multiple random 
sources in the upper convection zone. In time-distance helioseismology,
coherent wave signals are constructed by calculating a
cross-correlation function of oscillations observed at locations,
$\mathbf{r_1}$ and $\mathbf{r_2}$, separated by distance $\Delta$
and for time lag $\tau$:
\begin{equation}
C(\Delta, \tau) = \int_0^T f(\mathbf{r_1}, t) f(\mathbf{r_2}, t +
\tau) \mathrm{d} t,
\end{equation}
where $f(\mathbf{r},t)$ is the oscillation signal at location
$\mathbf{r}$, $\Delta$ is distance between $\mathbf{r_1}$ and
$\mathbf{r_2}$, and $T$ is the duration of the whole sequence.
 The cross-correlation for each distance
is obtained by averaging numerous cross-correlations calculated
between pairs of pixels, which are a given distance, $\Delta$, apart.

The time-distance diagram, as shown in Figure~2 in Paper I, and many
other time-distance diagrams published in literature, are actually a
collection of computed cross-correlations with continuous time lags,
displayed as a function of acoustic wave travel distance.  According
to the conjecture of \citet{ric00}, the
cross-covariance function may be considered as a signal from a point
surface source of some particular spectral properties.  Nearly all
the previously published time-distance diagrams were obtained at the
photospheric level, as no observations or simulations below the
photosphere were available before.

With the availability of 3D convection simulation, we are now
capable of computing the time-distance diagrams at different depths
beneath the photosphere by cross-correlating acoustic signals at a
desired depth with signals in the photosphere, and following the equation:
\begin{equation}
C(\Delta, \tau, d) = \int_0^T f(\mathbf{r_1}, 0, t) f(\mathbf{r_2}, d,
t + \tau) \mathrm{d} t,
\end{equation}
where $f(\mathbf{r}, d, t)$ is the oscillation signal at 
horizontal location $\mathbf{r}$, and at the depth of $d$ and time $t$.
Because of the great computational burden, in practice such 
computations are performed in the Fourier domain. The time-distance
diagram at a depth, $d$, shows the time it takes the acoustic wave
to travel from a point at the surface to a point at the depth $d$
and a horizontal distance $\Delta$ away.

\begin{figure}[!hb]
\epsscale{0.9} \plotone{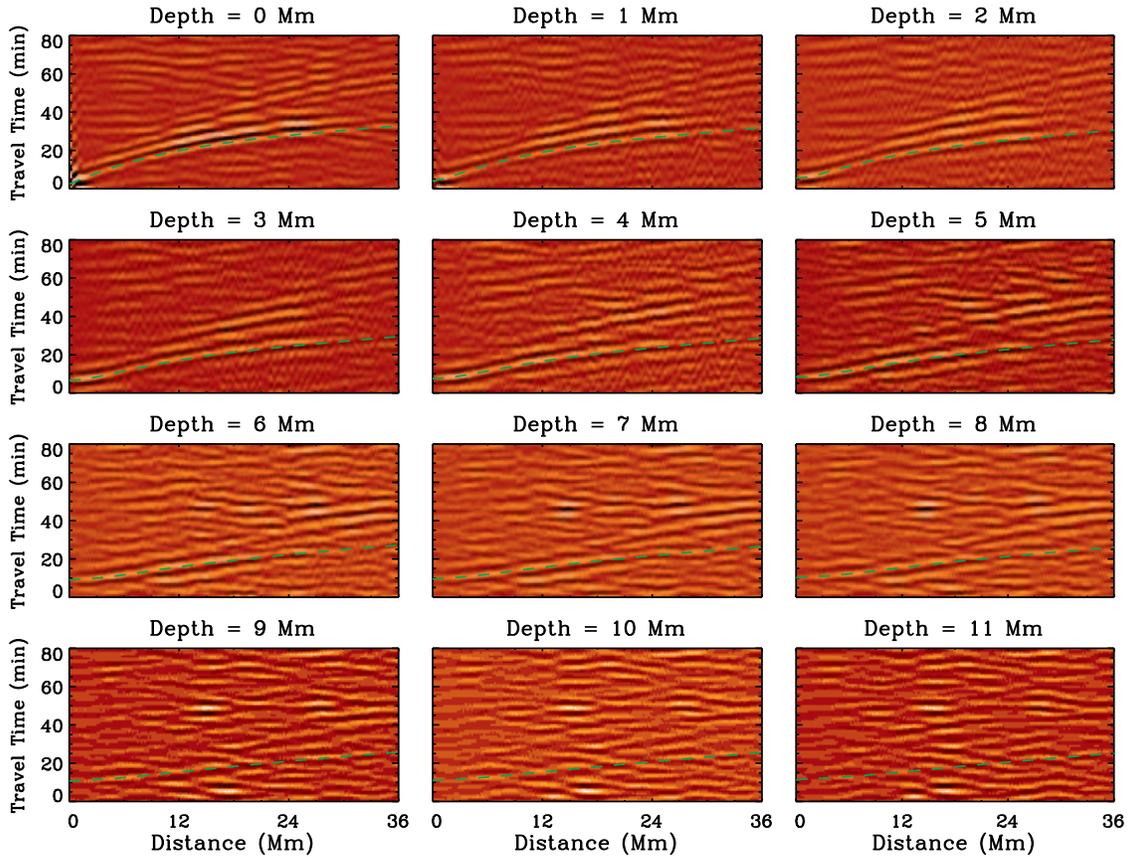} \caption{Time-distance diagrams,
$C(\Delta, \tau, d)$ calculated from the simulated data at several
selected depths, showing the ridges corresponding to the
relationships between acoustic travel times and distances at
different depths after an acoustic wave is initiated at the depth of
0 Mm. Green dashed lines in each diagram indicate expectations of 
the time-distance relationships based on the ray theory. } \label{fg2}
\end{figure}

We have computed nearly 100 such time-distance diagrams at every fifth 
depth of the simulated data, since the original data have 500 vertical 
mesh points, but some of these mesh points are above the photosphere 
and not used in computing these diagrams. Several selected depths are
presented in Figure~\ref{fg2}. The diagram at depth of 0 Mm is
actually the time-distance diagram for the photosphere. The
evolution of acoustic signals with time, and the propagation along
horizontal distance are clearly seen from 0 Mm to a depth of 8 Mm,
approximately. The time-distance diagrams match very well with 
the expected time-distance relationships derived from the ray theory,
as indicated by the dashed lines in each diagram. Below 8 Mm, 
the signals are not so clear as above.

\begin{figure}[!ht]
\epsscale{0.9} \plotone{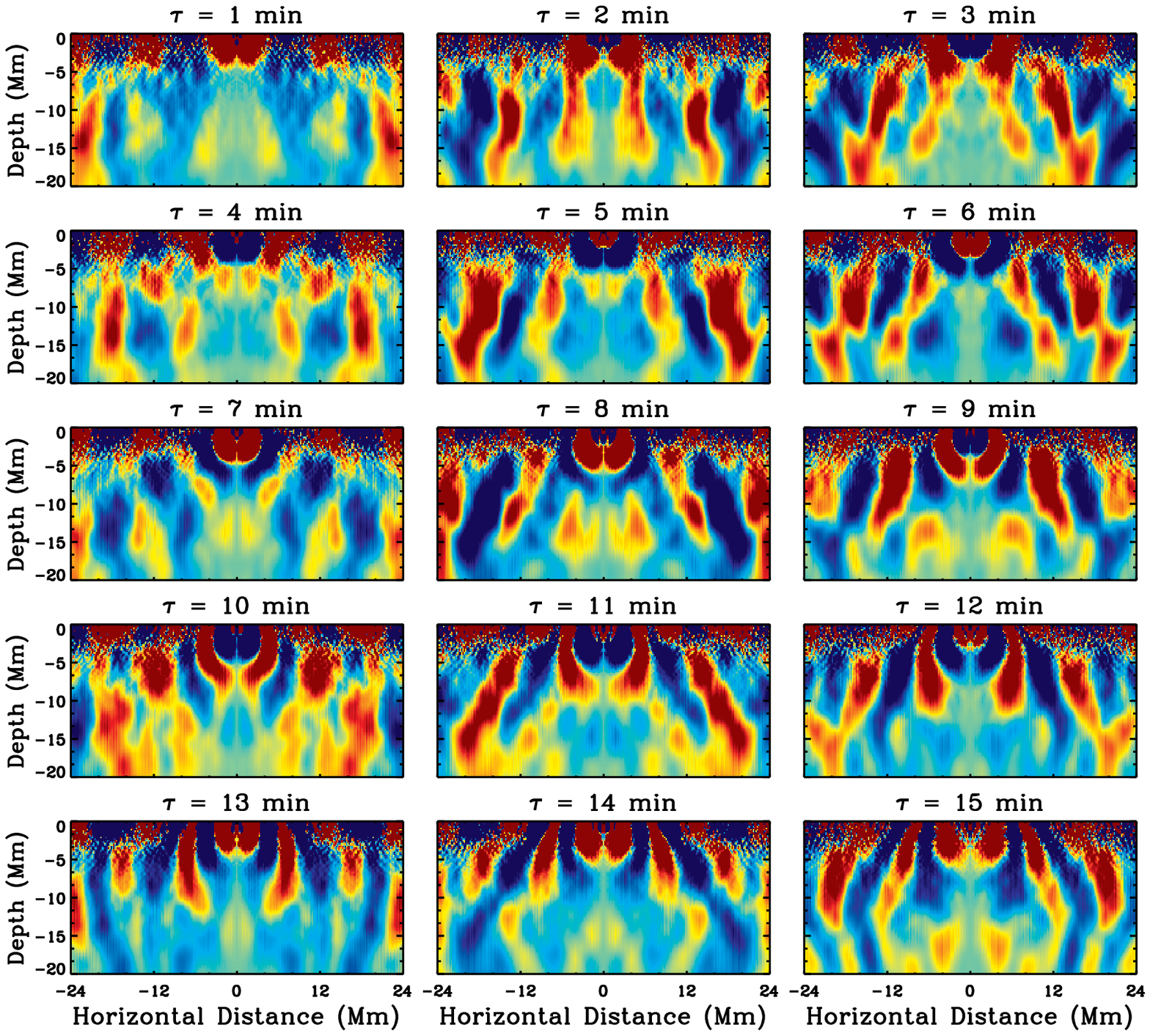} \caption{Selected snapshots showing
the propagation of acoustic waves into the interior, and reflecting
back to the surface. See the text for details.} \label{fg3}
\end{figure}

The two-dimensional time-distance diagrams for different depths can
be represented as a three-dimensional datacube with two spatial
dimensions, horizontal distance $\Delta$ and depth $d$, and time
$\tau$. It is interesting to display the spatial images as a time
sequence, showing acoustic waves propagating from the
surface into the interior. Some selected images of this 
sequence are displayed in
Figure~\ref{fg3}. The left side of each image is symmetrized with
the right side. In order to better show the weak signals in the
deeper regions, the signals near the surface are intentionally saturated.

As can be seen from the time-distance diagram at the depth of 0 Mm
in Figure~\ref{fg2}, the acoustic waves travel like a wave packet,
with an oscillation period of 5 minutes or so, and a width of about 17 
minutes \citep[also see][]{duv97}. In Figure~\ref{fg3}, at 
$\tau = 1$ min, one can see a very small blue feature at the top,
and this is considered to be the first negative oscillation in the wave packet. 
With the evolution of time the blue feature,
i.e., the wavefront, expands in size, and propagates into the
interior. It is followed by a signal of the opposite sign, a
red feature, which is initiated at $\tau=5$ min.
We noted that at some moments of the evolution, the
wave fronts become open at their horizontal central region. In the
upper 10 Mm or so, the general circular wave shapes are often kept
well, but below 10 Mm, the wave structure is often irregular and
noisy. Also, it looks like some signals, though weak, are
reflected back from the bottom boundary.

It is curious that the central part of the waves is open. This may
come from the following reason:
Because the horizontal span of the simulation domain is
only 48 Mm, the simulated data cannot carry any acoustic wave modes
that have a first bounce travel distance longer than 48 Mm. Clearly,
the modes which travel a longer first bounce distance contribute
most of the signal at the central part of the wavefronts. A simple
estimate of the turning depth of the acoustic waves corresponding to
a first bounce travel distance of 48 Mm is approximately 15 Mm. This
is roughly in agreement with the images in Figure~\ref{fg3}, in
which no clear signals deeper than 15 Mm can be seen. 

Despite the curious open structures at the central part of
wavefronts, the noise outside the wave features and some reflected
signals from the bottom, it is quite clear the waves are
substantial, clear, and in nice agreement with the theoretical
expectations. The constructed wave propagation clearly shows
that the waves are refracted back to the surface from different
turning points as expected from linear helioseismology theories. It
is a great success of the time-distance technique that it is capable
of retrieving acoustic wave propagation in the deep, very turbulent
interior, and it is also a great success for the simulations that
they reproduce the basic properties of acoustic waves in the solar
interior. Therefore, we conclude that this simulation dataset is
quite suitable for testing the $p$-mode time-distance analysis,
while keeping in mind that the deepest acoustic wave probe depth in
these simulations is shallower than 15 Mm.

\section{Propagation of Surface Gravity Waves: Scattering and 
Sensitivity Kernels}

The sensitivity functions (or kernels) represent perturbations of 
travel times to small localized features on the Sun. 
To evaluate the potential for using surface gravity waves (f-modes)
in the simulations for testing the $f$-mode diagnostics, we have
calculated empirical sensitivity functions following the recent work
of  \citet{duv06}. Their work has offered us not only another method
for checking the helioseismic reasonableness of the simulated data,
but also a meaningful approach to compute the sensitivity kernels
that is potentially useful for time-distance inversions. 

\citet{duv06} computed the sensitivity kernels for $f$-modes by 
measuring travel time variations around some small
magnetic elements from real MDI observations. They found elliptic
and hyperbolic structures in the kernels, which are similar to the
structures modeled by \citet{giz02} and attributed to wave
scattering effects. Because of wave scattering, the travel times are
sensitive not only to perturbations in the region of wave
propagation between two points but also in some areas away the way
path, along a hyperbolic-shaped regions. They concluded
that their measurements demonstrated that the Born approximation was
suitable for deriving time-distance inversion kernels, and that the
wave scattering effect is important and has to be included in the
derivation of the sensitivity kernels for helioseismic inversions.

Here, we try to check whether the sensitivity kernel with similar
hyperbolic structures can be obtained from the numerical
simulations, and thus  whether the wave scattering is properly
modeled in the simulated data.

Since there are no magnetic elements in the current simulation
dataset, we select areas with strong downdrafts as the features
to calculate the travel time sensitivity kernels. We average the
vertical velocity  over the whole time sequence of 511 minutes, and
select 50 small areas that have the strongest downward flows. We
set these 50 areas as the central features to compute the kernels.
We then follow the procedure described as ``feature method'' in
\citet{duv06}, and compute travel time variations around these
selected downdrafts areas. We follow every step described in this
paper, except that we use a longer time series in our computations,
and that we use the Gabor wavelet fitting to measure the wave travel
times. Although we have only one data sequence, and just 50 selected
downdraft areas available for averaging (unlike a number of
observational datasets and thousands of magnetic elements in the
\citet{duv06} paper), we were able to obtain the sensitivity kernels
with a reasonable signal-to-noise ratio. Still, we had to do
additional spatial averaging to recover the wave scattering
structures.

\begin{figure}[!ht]
\epsscale{0.6}
\plotone{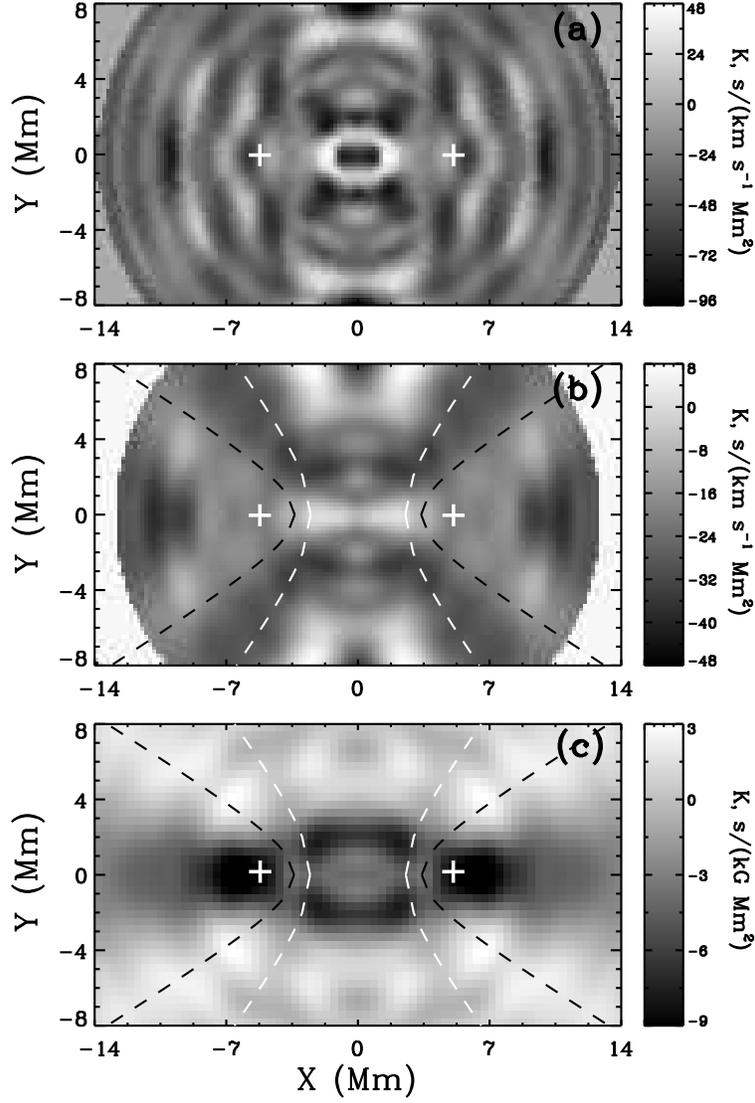}
\caption{(a) Sensitivity kernel measured around downdraft locations
from the simulated data using $f$-modes. (b) Sensitivity kernel obtained
by smoothing the kernel in (a). (c) Sensitivity kernel measured around
small magnetic elements from SOHO/MDI observations \citep[adapted
from][]{duv06}. The dark and light dashed hyperbolas in (b) and (c) 
indicate the locations of two hyperbolic structures observed in
(c). The white crosses in each panel indicate the locations of
observation points.}
\label{fg4}
\end{figure}

Figure~\ref{fg4} presents the results of our measurements a) before
and b) after boxcar smoothing. There are some small scale black and
white patches in the measured original travel time variations, and
large scale structures are not quite clear, although such structures
do exist. Such small scale structures are not seen in the kernels
obtained from real observations, and this may be possibly caused by 
the following reason: the numerical simulation data have a much
higher spatial resolution than the real observations, and thus our
measurements could pick up some small-scale signals that are not
resolved in the observational data. Additionally, the convective 
structures we use for such measurements are much less stable than 
the real magnetic elements, and this may also contribute some noises to 
the measurements.

To get a better signal-to-noise ratio for large-scale structures, we
applied a boxcar smoothing to the original measurements, and obtain
the result shown in Figure~\ref{fg4}b. The hyperbolic dark and light
features appear quite clearly, similar to the sensitivity kernel
measured from magnetic elements of real observations, although the
locations have slight offset as indicated by the dashed lines.  
However, the details are not comparable, because our sensitivity kernel is
measured around large downdrafts areas, but not for magnetic
elements as in Figure~\ref{fg4}c.

It is important that the hyperbolic structures are found in the
$f$-modes sensitivity kernels obtained from the simulated data,
because this demonstrates that the wave scattering effect is
reproduced in the numerical simulation of convection. However, at
present, it is difficult to use the empirical sensitivity kernels in
time-distance inversions, because the measured kernels such as the
one shown in Figure~\ref{fg4} may depend on various factors at the
downdraft locations, for instance, not only the vertical
component of flow, but also the converging flows that are often
associated with the downdrafts. This issue requires further
investigation.

\section{Time-Distance Helioseismology Test for Sub-Surface Flows}

\begin{figure}[!ht]
\epsscale{0.7}
\plotone{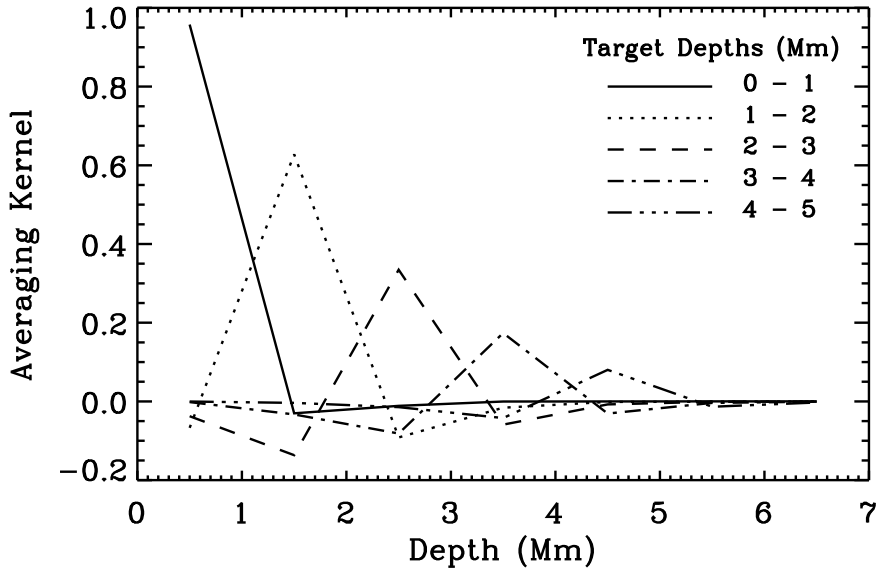} 
\caption{Averaging kernels for the selected target depths. Note that
each point is plotted at the middle of its corresponding depth interval.}
\label{fg5}
\end{figure}

To test the current time-distance helioseismology procedure, we
measured travel times of acoustic waves using only the vertical
velocity at the photospheric level from the simulated data, then
inferred the three dimensional velocities in the interior by using
the inversion procedure based on a ray approximation
\citep{kos97,kos00,zha01}, and finally compared the inversion
results with the averaged interior velocities from the simulations.
Following the typical $p$-mode time-distance measurement schemes
\citep[e.g.,][]{zha01}, we select the following seven annulus radii
to perform our measurements: 8.64 -- 10.37, 10.56 -- 12.29, 12.48 --
14.21, 14.40 -- 16.13, 16.32 -- 18.05, 18.24 -- 19.97, 20.16 --
21.89 Mm. The greatest annulus is a little smaller than half of the
horizontal span of the simulated data. In order to evaluate the
previous inversion results, the inversions here employ the ray
approximation kernels.

\begin{figure}[!ht]
\epsscale{0.7} \plotone{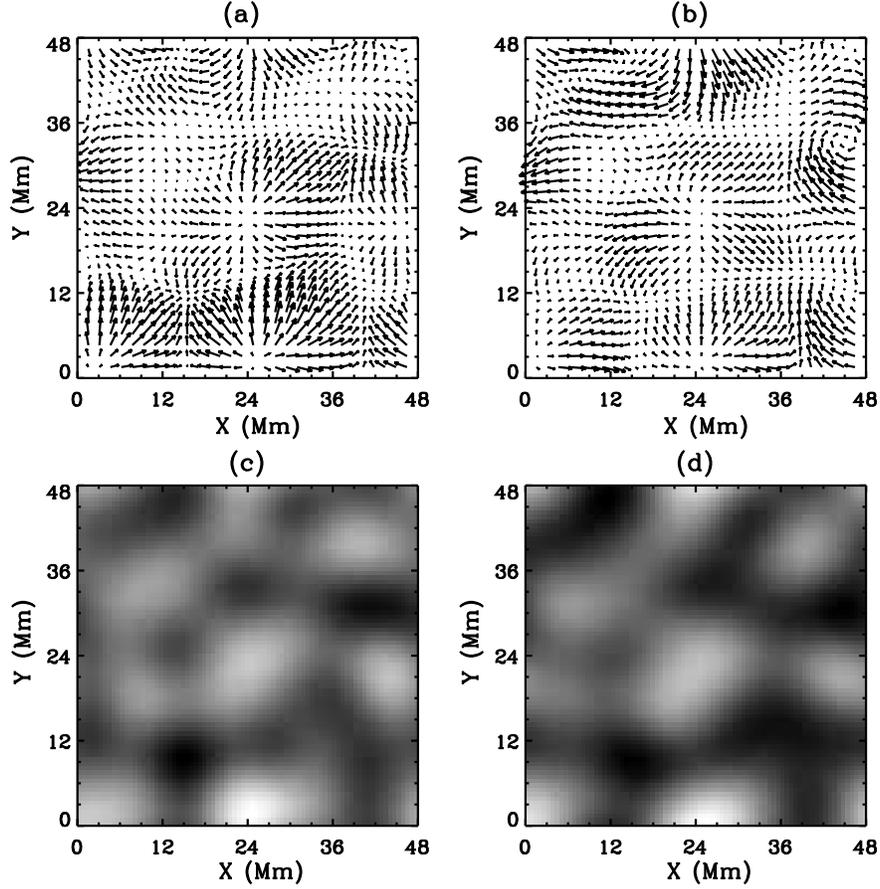} \caption{Comparison of the inverted
horizontal flow fields with the simulated data, at the depth of 1 -- 2
Mm: horizontal flow fields from the simulations (a) and the inversions (b),
and a divergence map from simulation (c) and inversion (d). Both
vector flow fields, and both divergence maps are displayed with same
scales. The longest arrow corresponds to 300 m/s.} \label{fg6}
\end{figure}

To evaluate the inverted time-distance results, we compare our
inversion results for flows with the actual flow fields directly
from the simulated data. The inverted results are often given as
averages of some depth ranges, for instance, 1 -- 2 Mm in
Figure~\ref{fg6}. Hence, the simulated data are also averaged 
arithmetically in the same depth interval over the 511 minutes.
For each target depth, the averaging kernels 
from the inversion are three dimensional, and Figure~\ref{fg5}
only presents a one dimensional curve corresponding to the central
point. In addition to the direct comparisons, it is interesting to 
convolve the three dimensional averaging kernels with the three 
dimensional simulated velocities, and see how the resultant velocities 
compare with the inverted velocities. In the following, we present
results from the direct comparison, but give the correlation
coefficients of both comparisons in Table~\ref{tb1}.

\subsection{Horizontal Flow Fields}

\begin{figure}[!ht]
\epsscale{0.7}
\plotone{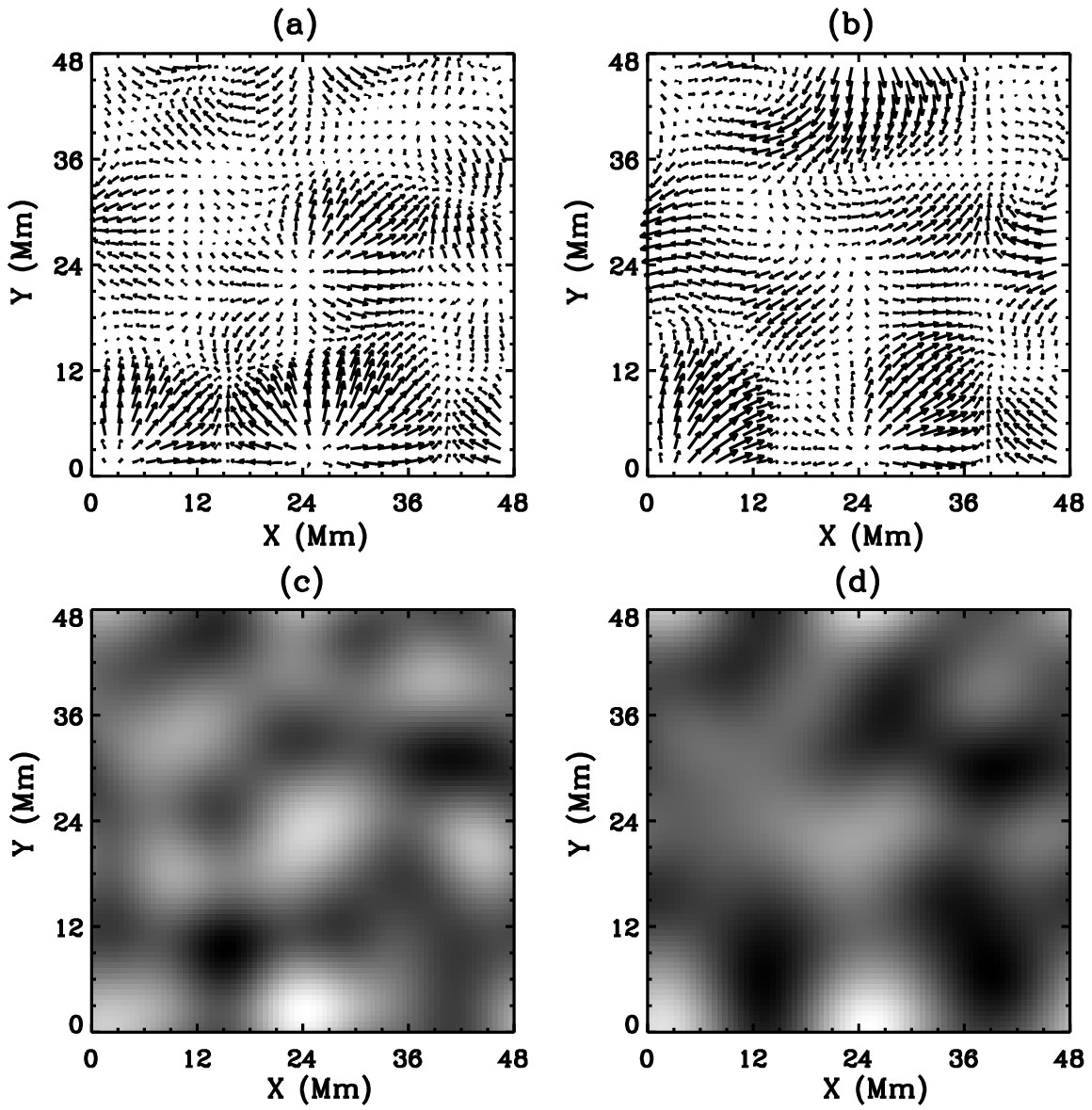}
\caption{Same as Figure~\ref{fg6}, but at the depth of 2 -- 3 Mm.}
\label{fg7}
\end{figure}

\begin{figure}[!ht]
\epsscale{0.7}
\plotone{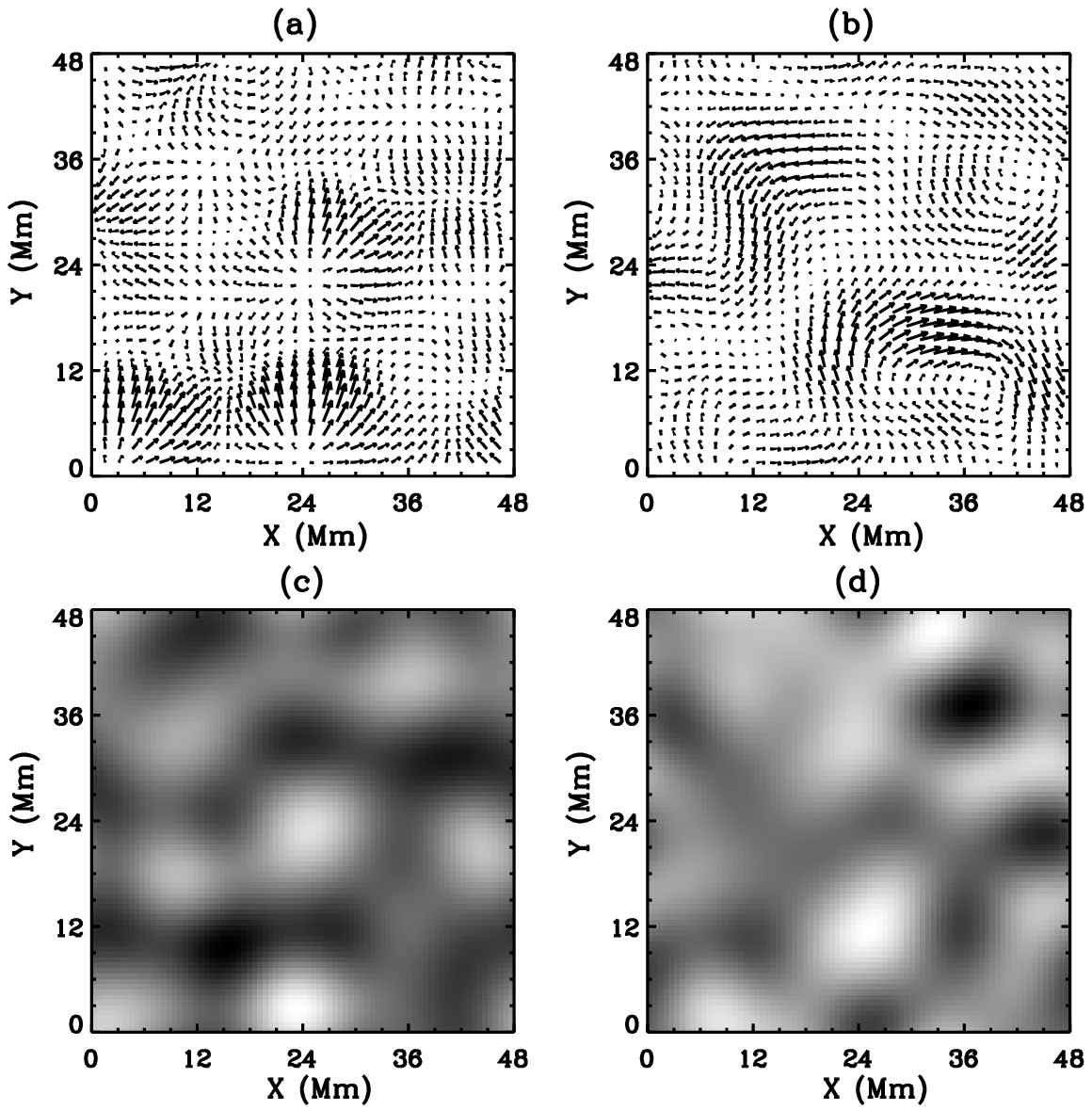}
\caption{Same as Figure~\ref{fg6}, but at the depth of 4 -- 5 Mm.}
\label{fg8}
\end{figure}

Figures~\ref{fg6}, \ref{fg7}, and \ref{fg8} present the inversion
results for the horizontal flow components for in three layers: 1-2
Mm deep, 2-3 Mm, and 4-5 Mm. Comparing the horizontal vector flows,
as well as the divergence map computed from the horizontal flows, we
find that at the depth of 1 -- 2 Mm the inversion results agree quite 
well with the time averaged simulation results. The results reveal areas of
strong flow divergence, which have a typical size of solar
supergranulations. This is in good agreement with the results
obtained by the time-distance analysis of $f$-modes and by a correlation 
tracking method, which are presented in Paper I.  These divergent 
areas are nearly in one-to-one correspondence between 
the inverted results and the
simulated data, with similar magnitudes as well. It should be
pointed out that the flow maps for the simulated data are displayed
after applying a low-pass filtering to only keep structures that
have a wavenumber smaller than 0.06 Mm$^{-1}$, in order to better match the
time-distance computational procedures, which undergo filtering and
smoothing during measurements and inversions.

As the inverted area deepens, the correlation between the inverted
results and the simulated data gradually worsens. At the depth of 2
-- 3 Mm one can still see those divergent flow patterns in both
data, but clearly not as clearly as at the depth of 1 -- 2 Mm. However,
at the depth of 4 -- 5 Mm and deeper, the inverted horizontal flows
show no clear correlation with the simulated data.

\begin{table}[!ht]
\begin{center}
\caption{Correlation coefficients between inverted results and
simulated data at different depths. The numbers shown in parenthesis
are correlation coefficients after the simulated data are convolved
with the inversion averaging kernels.}
\label{tb1}
\begin{tabular}{ccccc}
\tableline \tableline
depth & $v_x$ & $v_y$ & divergence & $v_z$ \\
\tableline
0 -- 1 Mm & 0.72 (0.72) & 0.64 (0.64) & 0.56 (0.56) & -0.72 \\
1 -- 2 Mm & 0.85 (0.87) & 0.76 (0.76) & 0.89 (0.91) & -0.72 \\
2 -- 3 Mm & 0.87 (0.92) & 0.74 (0.83) & 0.78 (0.84) & -0.29 \\
3 -- 4 Mm & 0.74 (0.84) & 0.37 (0.53) & 0.36 (0.50) & 0.34 \\
4 -- 5 Mm & 0.18 (0.51) & 0.25 (0.62) & -0.35 (-0.11) & 0.32 \\
\tableline
\end{tabular}
\end{center}
\end{table}

Table~\ref{tb1} presents the correlation coefficients between the
inverted results and the simulated data at different depths in two
horizontal velocity components, the divergence that are computed
from the horizontal components, and the vertical velocity (see the
next section), separately. It is clear that the shallow regions
often have better correlations than the deeper regions, except for
the vertical velocity. It is curious but not clear why the two
horizontal components have correlation coefficients that differ so
much, with the north-south direction often worse than the east-west
direction. A similar east-west and north-south asymmetry in time-distance
results was also found when comparing with local correlation tracking 
results \citep{sva06}. And, the correlation of the divergence maps 
also differ from that of the horizontal components. It is worthwhile
pointing out that after the simulated data are convolved with the 
inversion averaging kernels, the correlation coefficients are generally
improved, significantly in the deeper areas.

\subsection{Vertical Flow Fields}

\begin{figure}[!t]
\epsscale{0.8} \plotone{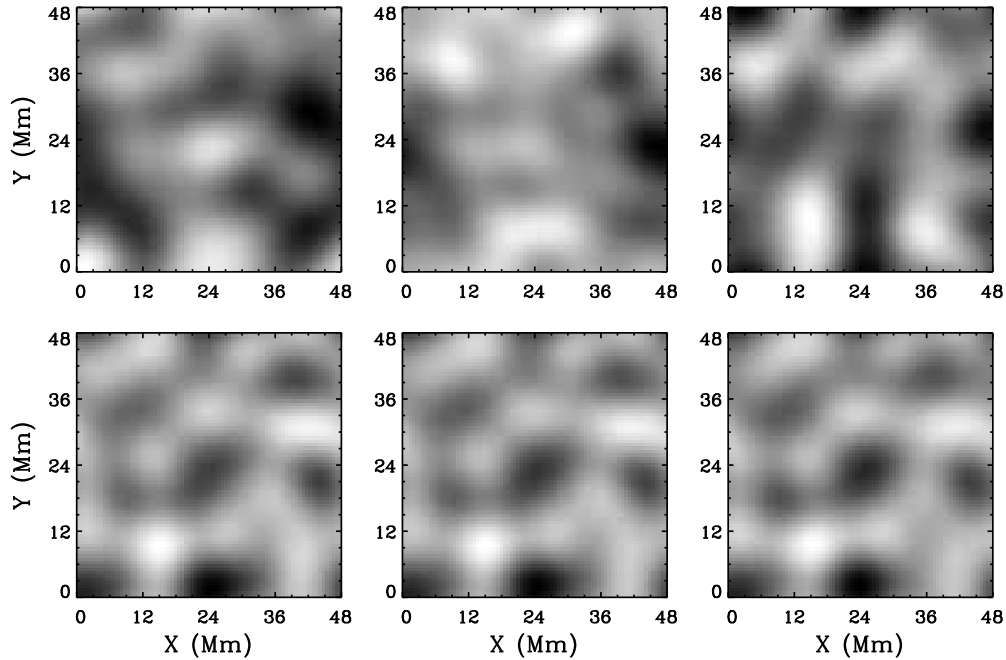} \caption{Comparison of the inverted
vertical flows with the simulated data. The upper panels shows the vertical
velocities from inversion at the target depths of, from left to right, 1 -- 2
Mm, 2 -- 3 Mm, and 4 -- 5 Mm. The lower panel shows the simulated
vertical flows averaged for the corresponding depth ranges. Dark
regions show upflows, and light regions show downflows. The velocity
range is from $-150$ to 150 m/s.} \label{fg9}
\end{figure}

It is often the case, in both real observations and numerical
simulations, that in the photosphere the vertical velocity is significantly 
smaller than the horizontal velocity of the same depth, due
to the strong density stratification in the upper solar convection
zone. And often, near the surface, the downward flows concentrate 
in very narrow lanes among boundaries of granules or supergranules. 
These properties make it more difficult to infer vertical velocities 
by the local helioseismology techniques, which often involves 
large area smoothing, because the small magnitude vertical
velocity in small areas can be easily smeared out.

Figure~\ref{fg9} presents results of time-distance inverted results
of vertical flows, along with the averaged vertical velocities from
simulated data at the corresponding depths. At the depth of 1 -- 2
Mm the inverted vertical velocities basically have opposite signs
to the simulated velocities. For the other two depths, no clear
correlation or anti-correlation is seen. As also presented in
Table~\ref{tb1}, the correlation coefficients show that, at shallow
depths, inverted vertical flows are in anti-correlation with the
simulated data, while at deeper depths, the correlations are
positive but weak. It seems that the inversions for the vertical
flow fields are completely unsuccessful. This discrepancy is not
only due to the small magnitude of vertical velocities, but is
also caused by the cross-talk effects discussed in the next section.

\section{Discussion}

The realistic simulation of solar convection obtained by
\citet{ben06} gives us an unprecedented opportunity to evaluate the
time-distance helioseismology technique and some other local
helioseismology approaches.

By computing the time-distance diagrams at different depths, as
shown in Figures~\ref{fg2} and \ref{fg3}, we have confirmed that
there are acoustic waves propagating in the interior in the simulated
data with properties similar to expected from the helioseismology
theory. Although there are some unexplained correlated signal
noises, bottom reflections, and open structure at the center of
wavefront, these do not affect our time-distance analysis in shallow
regions below the surface. The longest annulus used in our analysis
is about 21 Mm, which probes into a depth of approximately 6.5 Mm
according to ray theory, shallow enough not be affected by those
unexplained factors in the simulations. Our measurements of the
$f$-mode sensitivity kernel near the surface, shown in
Figure~\ref{fg3}, have demonstrated that the convection simulation
data have wave scattering properties similar to the real solar
observations. These analyses demonstrate that the simulated data have
the wave properties that are necessary for time-distance
helioseismology analysis.

Using very turbulent convection data at the photospheric level and
the time-distance technique based on a ray approximation, we were
able to derive the internal flow fields, which are in nice
correlation with the simulation results in shallow regions. It is
not surprising that the time-distance inversions cannot well resolve
properties in larger depths, which was already demonstrated by some
artificial data tests \citep[e.g.,][]{kos96, zha01}. It is believed
that the reliability of inversion results highly depends on the number
of ray-paths passing through that area, whereas deeper areas have
fewer ray-paths passing through and less information brought up to
the surface. Additionally, it should be recognized again that the
horizontal scale size of 48 Mm limits the deepest ray-path
penetration at a depth of approximately 15 Mm, and our longest
annulus radius once again limits our deepest probe to a depth of 6.5
Mm or so. Therefore, it is quite reasonable that our inversions give
acceptable results until a depth of only 4 Mm or so.

It is also not surprising to see the failure of vertical flows in
inversions. Certainly, the small magnitudes of vertical velocity may
be one reason. However, we believe that the main reason of the
failure is due to the cross-talk effects, as already demonstrated by
use of some artificial tests \citep{zha03}. The divergence inside
supergranules speeds up outgoing acoustic waves by the same way as
downdrafts do, and similarly, the convergence at boundaries of
supergranules slows down outgoing waves from this region similar to
what upflows do. The time-distance inversions cannot distinguish the
divergence (or convergence) from downward (or upward) flows,
especially when the vertical flow is small in magnitude. Although it
is believed that some additional constraints, e.g., mass
conservation, may help the inversion in resolving vertical flows,
current inversion technique of time-distance restricts itself using
pure helioseismological measurements.  This set
of simulated data may give us a very good test ground for the future
development of vertical velocity inversion codes.

Still, inversions in this study use the ray approximation kernels in
order to evaluate the old results published previously by use of
such kernels.  With the availability of Born approximation kernels
\citep{bir04}, it would be very interesting to test such kernels by
this kind of analysis on the current simulated dataset, although some
previous experiments \citep{cou04} showed that the inversion results
based on two different kernels did not differ much. It is expected
that the Born kernel may give better results in deeper areas, but
may not be able to solve the vertical velocity problem caused by the
cross-talk effects. New time-distance helioseismology schemes are
probably required for the solution.

The numerical simulation of solar convection give us an opportunity
to evaluate the time-distance technique in quiet solar regions.
However, it would be especially interesting if we could test this
helioseismology technique in a magnetized region, as the simulations
of magnetoconvection \citep[e.g.,][]{ste06, sch06} are extended in
both spatial and temporal scales to meet the helioseismological
measurement requirements. For instance, we can evaluate the accuracy
of the inferred sunspot structures and flow fields \citep{kos00, zha01}, 
and evaluate various magnetic field effects based on such numerical
simulations.

\acknowledgments

We thank Dr.~Tom Duvall for providing us data making Figure~\ref{fg4}c,
and Dr.~Takashi Sekii for insightful comments on interpretation of 
interior wave propagations. We also thank an anonymous referee for 
suggestions to improve the quality of this paper. The numerical 
simulations used in this work were made under support by NASA grants 
NNG04GB92G and NAG512450, NSF grants AST-0205500 and AST-0605738, 
and by grants from the Danish Center for Scientific Computing. 
The simulations were performed on the Columbia supercomputer of 
the NASA Advanced Supercomputing Division.

\end{document}